# Macro-scale Topology Optimization for Controlling Internal Shear Stress in a Porous Scaffold Bioreactor

K. Youssef[2], J.J. Mack[3], M.L. Iruela-Arispe[3], L.-S. Bouchard[1,2*]

Departments of Chemistry & Biochemistry[1], Biomedical Engineering[2] and Molecular, Cell and Developmental Biology[3], University of California, Los Angeles, CA 90095

*corresponding author: bouchard@chem.ucla.edu

**Abstract**

Shear stress is an important physical factor that regulates proliferation, migration and morphogenesis. In particular, the homeostasis of blood vessels is dependent on shear stress. To mimic this process *ex vivo*, efforts have been made to seed scaffolds with vascular and other cell types in the presence of growth factors and under pulsatile flow conditions. However, the resulting bioreactors lack information on shear stress and flow distributions within the scaffold. Consequently, it is difficult to interpret the effects of shear stress on cell function. Such knowledge would enable researchers to improve upon cell culture protocols. Recent work has focused on optimizing the microstructural parameters of the scaffold to fine tune the shear stress. In this study, we have adopted a different approach whereby flows are redirected throughout the bioreactor along channels patterned in the porous scaffold to yield shear stress distributions that are optimized for uniformity centered on a target value. A topology optimization algorithm coupled to computational fluid dynamics simulations was devised to this end. The channel topology in the porous scaffold was varied using a combination of genetic algorithm and fuzzy logic. The method is validated by experiments using magnetic resonance imaging (MRI) readouts of the flow field.

**Introduction**

Mechanical stimulation of cells has been an important area of research in regenerative medicine and tissue regeneration (Nollert *et al.*, 1991; Datta *et al.*, 2006; Stolberg and McCloskey, 2009). Several groups have focused on developing perfusion bioreactors that impose flow-induced shear stresses on cells residing within a porous matrix (Bancroft *et al.*, 2003; Dermenoudis and Missirlis, 2010; Brown *et al.*, 2008; Choi *et al.*, 2007). The research performed to date has demonstrated that shear stress is an important regulator of cell function and a relevant component for pre-conditioning cells prior to transplantation into live organisms (Reich and Frangos, 1991; Hillsley and Frangos, 1997; Smalt *et al.*, 1997; Klein-Nulend *et al.*, 1998; McAllister *et al.*, 2000; Jiang *et al.*, 2002). In previous work, local shear stresses were defined as a function of media flow rate and dynamic viscosity, bioreactor configuration and porous scaffold micro-architecture (Voronov *et al.*, 2010; Voronov *et al.*, 2010; VanGordon *et al.*, 2011). Target shear stress values were estimated as averages over the entire scaffold and changes in the overall average were achieved by varying the scaffold micro-architecture under similar flow rates (Porter *et al.*, 2005). Recent studies, however, have demonstrated the presence of complex internal microfluidic patterns within the architecture of the porous scaffolds that can alter cell growth in culture (Khoda *et al.*, 2011; Okkels *et al.*, 2011). Nonetheless, methods that convey precise shear stresses to cells that are, in addition, spatially distributed according to predetermined parameters are yet to be developed. Although modeling of scaffold architecture and flow conditions have been considered, no published work, to our knowledge, has systematically optimized the macro-architecture of channel topology within a porous scaffold to achieve a target shear stress under defined flow inputs. Such a method could have important advantages in that macroscopic flow conditions are easier to alter than the microstructural parameters, especially if real-time flow control is required.

In this article, we present a method to define channel topologies that yield a desired shear stress distribution. In our computational approach, channel topologies are governed by a genetic algorithm (GA) and shear stress distributions assessed by simulating flow using a Lattice Boltzmann Method (LBM). The GA achieves optimization by minimizing a cost function based on fuzzy logic rules. The optimized topologies demonstrate the ability to obtain a target shear stress with a more narrow distribution when compared to the case of a non-optimized topology. The method could be adapted to generate any desired distribution of internal shear stresses by modification of the cost function. To validate the optimization results experimentally, the trend in shear stress distributions under real flow conditions in porous scaffolds were derived from experimental measurements of flow assisted by magnetic resonance imaging (MRI) techniques. MRI is suitable for imaging flows non-invasively in optically opaque media (such as porous polymer scaffolds). This article provides proof-of-principle for the algorithm in 2D and its experimental realization by approximation using stacks of thin slices of scaffold material. Our method could be extended to optimizing flows by patterning channels at the microfluidic level, using micro-computed tomography (µ-CT) images as inputs to the pore geometry, and using 3D prototyping techniques for the structural reconstitution of the scaffolds. The MRI technique has the advantage that it can measure flows non-invasively in real-time, and should cell-seeded scaffolds be used, it may be possible to devise adaptive schemes to adjust the flow patterns as needed to account for the effects of cell growth.

## I. Materials and Methods

### 1.1 Lattice Boltzmann Method (LBM)

To model flows through porous media, a computational fluid dynamics (CFD) program was developed based on LBM with no-slip boundary conditions (Succi, 2001). Pores are modeled as open spaces in the lattice within which the fluid is allowed to flow. On the other hand, the walls surrounding a pore are modeled as closed spaces representing a solid, such that the fluid bounces back upon collision. A D2Q9 Bhatnagar-Gross-Krook (BGK) scheme was adapted to satisfy the evolution equation of the density distribution function and the conditions defined by Equations (1-3)

$$\Delta_i f_i = -\omega(f_i - f_i^e) \qquad (1)$$

$$\Sigma_i f_i^e = \Sigma_i f_i = \rho \qquad (2)$$

$$\Sigma_i f_i^e c_{ia} = \Sigma_i f_i c_{ia} = \rho u_a \qquad (3)$$

where $f_i$ is the density distribution function, $f_i^e$ is the equilibrium density distribution function, $\rho$ is the density, $a = x, y$ and $u = (u_x, u_y)$ as velocity, $c$ the velocity direction and $\omega = 1/t$ with $t$ as the time scale. These conditions lead to a generic family of LBGK equilibria that are expressed by Equation (4) in which $Q_i$ is the projector along the $i^{th}$ discrete direction. For the D2Q9 scheme, values for the weights $w_i$ and the speed of sound $c_s$ are fixed according to Equation (5).

$$f_i^e = \rho w_i \left(1 + \frac{c_{ia} u_a}{c_s^2} + \frac{Q_{iab} u_a u_b}{2 c_s^4}\right) \qquad (4)$$

$$c_s^2 = \frac{1}{3}; \ w_9 = \frac{4}{9}; \ w_{1,2,3,4} = \frac{1}{9}; \ w_{5,6,7,8} = \frac{1}{36} \qquad (5)$$

Shear stress was calculated under the Newtonian fluid assumption using Equation (6)

$$\overleftrightarrow{\tau} = v\left(\frac{\nabla U + \nabla U^T}{2}\right) \qquad (6)$$

where $\overleftrightarrow{\tau}$ is the shear stress tensor, $v$ is the dynamic viscosity and $U$ is the 2D velocity vector field. Shear stress $\tau$ was calculated as the largest eigenvalue of the tensor $\overleftrightarrow{\tau}$. The derivatives of the velocity field for the gradient were calculated by finite-difference approximation for the four partial derivatives at each point on the lattice corresponding to the fluid phase. The finite-difference approximations are given in Equations (7-10):

$$\frac{\partial U_x(i,j)}{\partial x} \approx \frac{U_x(i+1,j) - U_x(i-1,j)}{2} \qquad (7)$$

$$\frac{\partial U_x(i,j)}{\partial y} \approx \frac{U_x(i,j+1) - U_x(i,j-1)}{2} \qquad (8)$$

$$\frac{\partial U_y(i,j)}{\partial x} \approx \frac{U_y(i+1,j) - U_y(i-1,j)}{2} \qquad (9)$$

$$\frac{\partial U_y(i,j)}{\partial y} \approx \frac{U_y(i,j+1) - U_y(i,j-1)}{2} \qquad (10)$$

Simulations were performed on a 128×128 lattice and the method was validated against "textbook" problems with known analytical solutions: for laminar flows, bench mark models such as the parallel-plate model were used (the simulation results were compared with the analytical solutions for flow between two infinite plates); for turbulent flows, vortex streets around a cylinder were simulated.

### 1.2 Generation of the Channel Topology

To generate channel topologies, we create a set of curves, each of which is represented by a $(L/2)$-dimensional vector where $L$ is the length of the flow chamber (length is measured along the principal direction of flow). This vector is obtained by calculating the inverse discrete cosine transform (IDCT) of a second vector of an equal dimension. The second vector is zero-filled, except for a window near the start of the vector, and these non-zero entries determine the spatial frequency content of the curve. The number of discrete cosine components used determines the complexity of the curve, which is reflected by the maximum number of turns. A large number of components, however, increases the parameter space for the topology search and channel complexity and could potentially lead to long optimization times and difficulty in patterning the channels.

A set of curves is obtained, such as illustrated in Figures 1(A) and 1(B). These curves are superimposed, as shown in Figure 1(C), forming one quadrant of the 128×128 lattice. They are then symmetrically replicated to cover the entire lattice [see Figure 1(D)]. Finally, image processing techniques are used to obtain the desired channel width and the final topology as shown in Figure 1(E).

The DCT is known for its energy compaction property, whereby signal information tends to be concentrated in the lower frequency components of Fourier space. Using a topology generation technique that is based on the superposition of curves, we take advantage of the DCT compaction property to greatly reduce the number of parameters in the search space. This makes the optimization more tractable by reducing the search time. This approach also offers additional advantages, such as the ease of manufacturing and simulation. In principle, however, it should be possible to realize more complex scaffold geometries using 3D prototyping printers. In this case, one could envisage using entirely different methods for generating the topology.

**1.3 Topology Optimization by Genetic Algorithm (GA)**

The GA finds an optimal channel configuration under a set of fixed parameters, i.e. porosity, channel width, number of channels and flow rate. We note that the flow of fluid requires a connected pore space. This can be achieved by considering only pore spaces that are connected; however, doing so would require a detailed simulation of the 3D pore space geometry, leading to substantial increases in computational requirements. In our case, we have instead simulated the pore space using an "effective-medium" approach and our way of ensuring a connected pore space is to fix the porosity to a sufficiently large value. The search is performed in an evolutionary manner for a set of discrete cosine components that generate a channel topology using the mechanism described in Section 2.2 to obtain an optimal channel geometry that yields a uniform target shear stress throughout the porous media.

The search begins with a population of random solutions in the search space, which is defined by all combinations of non-zero entries of each vector. It then converges towards an optimal solution by evaluating the performance of each member in the population and passing the best two members to the next generation in a survival-of-the-fittest manner. The genes of the members with the best performance are crossed over and a random mutation is performed. The resulting offsprings are then added to the population as new members. As the generation number increases, members with improved performance evolve. An illustration of the evolutionary process carried out by the GA is shown in Figure 2 where the shear stress histograms and distributions for members in subsequent generations are shown. Each member in the population represents a channel topology and the performance evaluated by simulating fluid flow using the LBM to determine shear stress distribution and assign a score based on a cost function. The ultimate goal is to find the channel topology that leads to a minimum of the cost function.

### 1.4 Cost Function

The L-1 norm error, $\varepsilon$, can be used to measure the difference between the shear stress distribution $\tau(i,j)$ of a given channel topology and a target distribution $\gamma$ as described in Equation (11)

$$\varepsilon = \sum_{(i,j)\in \Lambda} |\tau(i,j) - \gamma| \qquad (11)$$

where $\Lambda$ is the set of lattice coordinates.

The goal of this optimization is to determine a channel topology that produces a target average shear stress that is uniformly distributed. The cost function can also be defined by the combination of the average and the standard deviation as in Equation (12)

$$\varepsilon = w_1 \cdot (\mu - \gamma) + w_2 \cdot \left(\frac{\delta}{\mu}\right) \qquad (12)$$

where $\mu$ is the average shear stress and $\delta$ is the standard deviation. Parameters $w_1$ and $w_2$ are weights that determine the influence of each component on the total error. This cost function is a linear relation that can be interpreted as a decision process that attempts to estimate the performance based on two separate values: $(\mu - \gamma)$ and $(\delta/\mu)$. For example, if $(\mu - \gamma)$ is deemed a more important factor in determining the error, it is assigned a larger weight.

Equation (12) is not the best cost function to use if we want to assign more weight to $(\delta/\mu)$ when its value is exceedingly low, which is usually the case when optimizing for a narrow distribution of shear stress. An interesting alternative is to use fuzzy logic to model the decision process. Fuzzy logic is a mathematical tool for dealing with uncertainty. It provides a mechanism for representing linguistic constructs such as "small," "medium," "good" and "few" to model the ambiguity associated with vagueness and imprecision (Sivanandam *et al.*, 2007). A possible fuzzy logic system is to evaluate the performance of a given channel topology based on a set of four rules: if $(\mu - \gamma)$ is *small* and $(\delta/\mu)$ is *small* then error is *small*; if $(\mu - \gamma)$ is *big* and $(\delta/\mu)$ is *small* then error is *medium*; if $(\mu - \gamma)$ is *small* and $(\delta/\mu)$ is *big* then error is *medium*; if $(ì - \gamma)$ is *big* and $(\delta/\mu)$ is *big* then error is *large*. Compared to the L-1 norm, this fuzzy logic method gives an improvement of ~15% in shear stress standard deviation when the average stress matched the target shear stress. A comparison of the relation between $(\mu - \gamma)$ and $(\delta/\mu)$ using Equation (12) versus using fuzzy logic is shown in Figure 3. Membership functions were represented by

Gaussian curves. The nonlinear relationship enables the GA to reach an optimal configuration more efficiently.

**2.5 Flow Experiments on Polymeric Scaffolds**

*Scaffold Fabrication:* Our theoretical results (to be described in Section 3) were validated in experiments on porous polymer scaffolds fabricated with flow channels of various topologies. Porous scaffolds were prepared by a porogen leaching method to achieve high porosity. Sugar crystals (250-355 µm size distribution) served as the porogen and were added to a 20 weight percent solution of polycaprolactone (PCL) in dichloromethane and thoroughly mixed to form a viscous paste (14:1, sugar:PCL). The sugar/PCL paste was then added to a Teflon mold (2 cm × 2 cm) machined with the desired channel topology as displayed in Figure 4(A). A Teflon plunger was applied to uniformly distribute the paste within the mold and compress the sugar crystals, thereby resulting in an interconnected network of pores upon removal of the sugar. Once the paste has been distributed and compressed, the scaffold was allowed to cure via solvent evaporation overnight. To achieve open porosity, the scaffold was leached in deionized water for several days, refreshing the water several times a day. To achieve a desired thickness, multiple scaffolds can be stacked as shown in Figure 4(B).

*Microstructural Characterization of Scaffolds:* X-ray micro-computed tomography (µ-CT) of the porous scaffolds revealed pore sizes on the order of ~250 µm; a representative microtomographic image is displayed in Figure 4(C). The µ-CT data was obtained on a SkyScan 1172 with 13µm spatial resolution and then analyzed using Matlab (Mathworks, Natick, MA) and iMorph softwares (http://imorph.sourceforge.net). The 3D data was used to compute the porosity (>70%), specific surface (20,164 $m^2/m^3$) and pore size distribution as displayed in Figure 4(D-E). These parameters (obtained in this study via µ-CT) should not be taken as true values upon which the success of any method hinges, as the actual values could be different, depending on the spatial resolution and imaging modality used to characterize the scaffold topography. For example, scanning electron microscopy (SEM) revealed more intricate pore features than is observed in the µ-CT scan [see Figure 4(F)].

*Bioreactor Construction:* To test the porous scaffolds under flow, a cylindrical bioreactor was custom designed and built. The bioreactor was machined from Teflon in two half-sections to house a 2 cm × 2 cm square and 1 cm-deep flow chamber, shown in Figure 4(G). Inlet and outlet ports were drilled through the Teflon to allow fluid to enter and exit the flow chamber. Plastic tubing was connected via Swagelok brass fittings to the inlet and outlet ports. To perform the flow experiments, the scaffold of interest was placed at the base of the flow chamber so that the inlet and outlet were centered at the scaffold surface. To fill the chamber, two non-channeled porous scaffolds were then added and compressed by tightening four nylon screws to seal the bioreactor.

*Flow imaging:* A 400 MHz vertical-bore Varian NMR system was used in all experiments. The bioreactor was placed in the bore of a 40mm-i.d. imaging probe. For flow imaging, a spin-echo multi-slice (SEMS) sequence was modified to perform phase-contrast MRI velocimetry (Callaghan, 1991) as follows. All existing gradients, except the phase encoding gradient, were flow compensated. The flow compensation (F.C.) gradients added are shown as green lobes in Figure 5(A). Pairs of flow weighting (F.W.), bipolar gradients were added along x, y and z axes to select the gradient first moment ($M_1$). Two experiments were first performed under flow with two gradient values, $+M_1$ and $-M_1$, where the value of $M_1$ was chosen large enough to include the highest anticipated flow velocity and avoid phase wrap-around effects. The two experiments were subtracted in order to obtain a velocity map. The velocity maps, however, contained artifacts from gradient non-idealities (e.g., eddy currents and non-linearities) and therefore the contribution from non-idealities was subtracted from an identical experiment performed in the absence of flow. With this approach, we were able to measure flow velocities below 1 mm/s using a total of 12 scans: 3 gradient directions ($x, y, z$) × 2 gradient reversals ($M_1$, $-M_1$) × 2 runs (flow, no flow). The imaging slice (1.0 mm thick) was centered on the inlet/outlet of the flow chamber, which fed into the patterned surface of the porous scaffold, as shown in Figure 5(B). The remaining imaging parameters were set to TR=5s, TE=50ms, 256×256 matrix and a field of view (FOV) equal to 25mm × 25mm. The source code for the Varian pulse sequence used in these experiments is available for download free of charge at our web site: `http://3Dneovascular.mcdb.ucla.edu`

## II. Results

### 2.1 LBM Simulations

LBM simulations were performed for three different scaffold configurations. The first configuration is a porous scaffold with no channels. The second configuration is a porous scaffold with four parallel channels connected to the flow chamber inlet and outlet. We will refer to this configuration as the "manual design." The third configuration is a porous scaffold with channel topology that was optimized using the topology optimization method. Since this study is a proof-of-concept for the optimization strategy, a number of constraints were imposed for time efficiency and to allow a fair comparison to the manual design topology. Accordingly, the optimization was performed by enforcing comparable channel width and number of channels. Although this limited the full potential of optimization, the resulting topology demonstrated a substantial improvement over the control (manual design). All configurations simulated a scaffold porosity of 95% (in 2D) and average flow velocity at the inlet of ~26 mm/sec. Based on results for random sphere packings in *n*-dimensions (Torquato, 2001), a 95% porosity in 2D likely corresponds to a lower porosity in 3D.

The simulation results of the three configurations are shown in Figure 6(A-C). The shear stress histograms are normalized by the standard deviation and the shear stress maps are plotted on a logarithmic scale to simplify comparison of the three channel topologies. When a target shear stress value of 2 (logarithmic scale) was set for the optimization, an improvement in shear stress uniformity and distribution is observed for the optimized configuration. Figure 6 shows the standard deviation divided by mean shear stress ($\delta/\mu$) for each configuration as a measurement of distribution uniformity. The optimized configuration was more uniform than both the no-channel (10.6%) and the manual design (5%). We note than 10% improvement in the quantity ($\delta/\mu$) is actually quite a large improvement in the shear stress uniformity: the value ($\delta/\mu$) optimized for is volume-averaged over the entire bioreactor including regions near the wall which see zero flow. It is, for example, impossible to improve on the values of this ratio near the reactor boundaries due to the no-slip condition. The best way to see the substantial improvement that can be achieved using our method is to compare the results of Figure 6 (bioreactor with no channels) to the optimization of Figure 8 presented later.

In all cases, simulations produced a nonlinear relationship between input pressure and average shear stress. For example, in the case of the no-channel configuration, a maximum input pressure was reached after which no further increases in the average shear stress were observed. The introduction of macro-scale channels to the scaffold alters the average shear stress limit and by implementing our optimization method, we were able to systematically vary the average shear stress limit to a desired target, whereas without topology optimization, this value remained below the plateau maximum. Furthermore, our optimization method offers flexibility to control different aspects of shear stress distribution by modifying the rules of the fuzzy logic cost function. In other designs (not shown here), the cost function was modified to assign more weight to minimize standard deviation thereby producing a channel topology with a value of ~0.7. This relative deviation can be further optimized by increasing the number of channels and reducing the width of the channels. However, pushing the level of complexity makes it more challenging to realize the patterned scaffold experimentally. This is where improvements in our method could be realized using a 3D prototyping printer instead of CNC milling.

### 2.2 Flow Rate Simulations

Varying the inlet flow rate varies the applied pressure difference across the flow chamber of the bioreactor. Simulations were performed for 10 different flow rate values that ranged from very low (0.1 mm/s) to very high (5 mm/s). Five simulations for each flow rate were performed with the same target shear stress and uniformity measured as the ratio between standard deviation and average shear stress. Figure 7 plots the relationship between shear stress uniformity and flow rate. Analyzing this relationship, the ideal flow rate for a target shear stress value can be determined and the overall performance (least fluctuations between runs) can be identified. Large fluctuations were observed for very low and very high flow rates. At very low flow rates, fluctuations are mainly due to diffusion. At very high flow rates, the fluctuations are likely due to an inability to find a channel topology that can match the high flow with the given target shear stress. In other words, if the flow rate is too high for the target shear stress, no channel topology can be found to match the target prescription. This is the main cause for the observed random fluctuations from one run to another.

## 2.3 Flow Rate Optimization

In the previous optimizations, the flow rate was kept fixed during the optimization. We also investigated the effects of flow rate on shear stress distribution when the flow rate is allowed to vary during the optimization. In this case, the GA searches for a combination of topology and flow rate to achieve an optimal distribution for a target shear stress. The resulting channel topology is displayed in Figure 8, where the flow rate corresponds to an inlet velocity value of 91 mm/s. Improvement was observed both visually and quantitatively since a smaller $(\delta/\mu)$ value was obtained. This suggests that removal of the flow rate constraint gives the GA more freedom to optimize the channel design.

## 3.4 Flow Imaging Experiments on Fabricated Scaffolds

Each of the fabricated 2 cm × 2 cm porous scaffolds was placed into the bioreactor flow chamber and water was flowed at a constant rate of ~5 mL/min using a syringe pump (Harvard Apparatus). The flow chamber was filled using additional non-channeled porous scaffold material to minimize water flowing outside the channel region and making sure that the surface of each scaffold of interest [see Figures 9(A-C)] was aligned with the chamber's inlet and outlet. The 2D MRI slices were imported into MATLAB (The Mathworks, Natick, MA) for analysis. The flow velocity maps yielded three orthogonal components of the velocity, $\vec{v} = (v_x, v_y, v_z)$, at each point in space. A plot of the magnitude of the velocity, $|\vec{v}|$, is shown in Figures 9(D-F). Velocity fluctuations were estimated from data acquired at zero flow and yielded errors (1σ limits) on the order of 0.05 mm/s. From the velocity maps, shear rates were calculated by finite-differences in the plane of the slice (flow was in-plane) using Equation (13).

$$\dot{\gamma} = \frac{\partial_x v_y + \partial_y v_x}{2} \qquad (13)$$

The absolute values of the shear rates at each pixel were retained for analysis and the resulting shear rate maps are shown in Figure 9(G-I). Histograms of the shear rate distributions were plotted normalized to the number of values in Figure 9(J-L). We note that this analysis neglects the z-component of velocity, which could reach up to 30% of the velocity magnitude in some regions of the scaffold. The values of $(\delta/\mu)$ for the shear rate distribution are listed in the insets of Figure 9(J-L). Substantial improvement was observed for the manual design compared to the case of no channels. Likewise, a narrower distribution of shear rates is measured in the topology-optimized scaffold as compared to the manual design and the no-channel scaffold. This demonstrates our ability to control the variance of the shear rate distribution. We note that the shear rate maps (Figure 9 G-I) do not match exactly those of the simulations performed (Figure 6) and we speculate that the divergence is likely due to the fact that the experimental flow patterns are inherently 3D whereas the shear rate analysis was performed using only two of the velocity components. True 3D analysis of the flow field could be possible by adapting our MRI flow sequence to perform 3D tomography instead of 2D slice readouts. Nonetheless, it is noteworthy that the trends in $(\delta/\mu)$ measured experimentally match those predicted theoretically and performance of the topology-optimized scaffold exceeded our theoretical predictions.

### III.    Discussion

All the simulations in this work are done in two dimensions (2D). The same method could be extended to control shear stress distributions in three dimensions (3D). The problem of computation time could be solved using parallel processing techniques. In our 2D calculations, the population size used in the GA is 20 and the number of generations is 100, for a total of 2,000 runs. Considering the size of the search space, the number of runs performed by the GA is quite low. For example, the number of parameters used by the GA in the flow rate simulations is

28. If we consider a resolution of $10^{-3}$ for each parameter, this produces a search space of $1,000^{28}$ possibilities. Using an average personal computer, each run takes an average of 9 seconds for a 128×128 lattice, or ~5 hours to perform a single optimization. To determine the computational time required for 3D optimizations using porosity maps from experimental data, we performed small-scale 3D LBM simulations of the flow field on the µ-CT data. The 3D LBM simulations showed a linear relationship in computational time with the 2D LBM simulations. This means that a 128×128×128 3D lattice requires 128 times the time required by a 128×128 2D lattice for equivalent time steps. Additionally, the convergence time will increase with lattice size and the 3D lattice will require more time steps to converge.

In preliminary tests conducted on a GPU array (nVIDIA 460GTX) the speed enhancements observed were as high as 28 times, compared to a 3.2GHz Intel CPU (non-parallel computation). The 460 GTX card contains 336 cores with 650MHz clock speed and 1GB memory. On a nVIDIA TESLA card containing 448 cores with 1.15 GHz clock speed and 3GB memory, speed enhancements of 50 times would be expected. Given the increase in computational time due to large lattice size, convergence time and speed enhancement from GPU implementation, an estimate for 3D simulations is approximately 20 times the duration of 2D simulations. Thus, a 128×128×128 3D optimization is expected to take approximately 4 days.

In future work, the biological response to the scaffold channel topology could be experimentally studied. Control over flow and shear stress within a 3D porous scaffold is essential in order to correlate cell response to a defined shear stress distribution. By controlling the scaffold macro-architecture, we can control flow in three dimensions and study the effects of shear stress, uniform or varied, on cell growth (i.e., cell proliferation and migration). The ability to vary the scaffold topology allows us to control patterns of shear stress within the scaffold. In this study, we focused on producing uniform patterns of shear stress. However, the spatio-temporal patterns of shear stress required for optimal cell growth and proliferation over time are presently unknown. Therefore, the ability to alter the flow pattern in both space and time will be paramount to understanding a biological response to flow and correlated mechanical forces. This need for adaptable flow patterns is especially relevant considering cell growth within a scaffold is expected to increase the resistance to flow. The MRI technique is particularly suited for use in adaptive control schemes during growth, due to its ability to provide real-time, non-invasive measurements. To control flow in 3D, multiple inlets can be arranged such that patterns of shear stress are applied to stimulate cell growth within the scaffold and ultimately improve ex-vivo growth of complex tissues in three dimensions. While we have not tested the effects of patterned flows on cell growth in this study, many research groups are pursuing such investigations. The high-porosity and large-pore-size PCL scaffolds used in our experiments feature open pores on the surface that facilitate the seeding of the scaffold via injection or gravity-mediated transport and diffusion. We note that any other scaffold material could potentially be used instead of PCL provided that it is permeable to fluid flow and the porosity is known to a reasonable degree.

## IV. Conclusion

In this paper we presented an algorithm to control shear stress distributions in porous polymeric scaffolds through optimization of macro-scale channel topology. Experimental results confirm the validity of our 2D optimization methods and future work will focus on extending these methods to control shear stress distribution in 3D. Parallel processing techniques are under consideration to reduce computation time and extend the optimization method to 3D geometries. Providing a means to control shear stress through macro-architecture of a scaffold could facilitate investigations of shear stress distribution on cell proliferation and gene expression under flow in a 3D scaffold. The main advantage of redirecting macroscopic flows is the ability to optimize the shear stresses throughout the entire bioreactor volume without the need to optimize the scaffold microstructure. The work could be extended to include 3D prototyping techniques for the production of scaffolds with more complex geometries. In this case, both microfluidic and macroscopic control of the flow topology may be achievable.

*Acknowledgments:* L.-S. B. thanks Emmanuel Brun for providing us with the iMorph software free of charge.

**Figures**

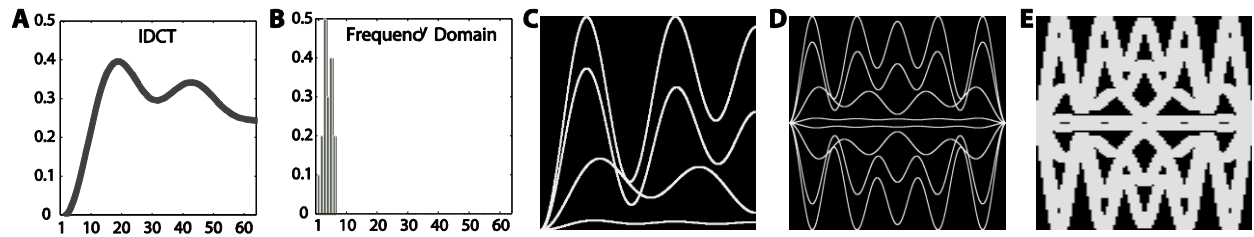

**Figure 1.** Illustration of the generation of a channel topology. (A) An initial curve is formed by the inverse discrete cosine transform (IDCT) of the vector of discrete cosine components shown in (B) where only the first six elements are allowed to take non-zero values. (C) Four curves are superimposed followed by (D) symmetrical replication (up/down, left/right) to yield the final topology, (E).

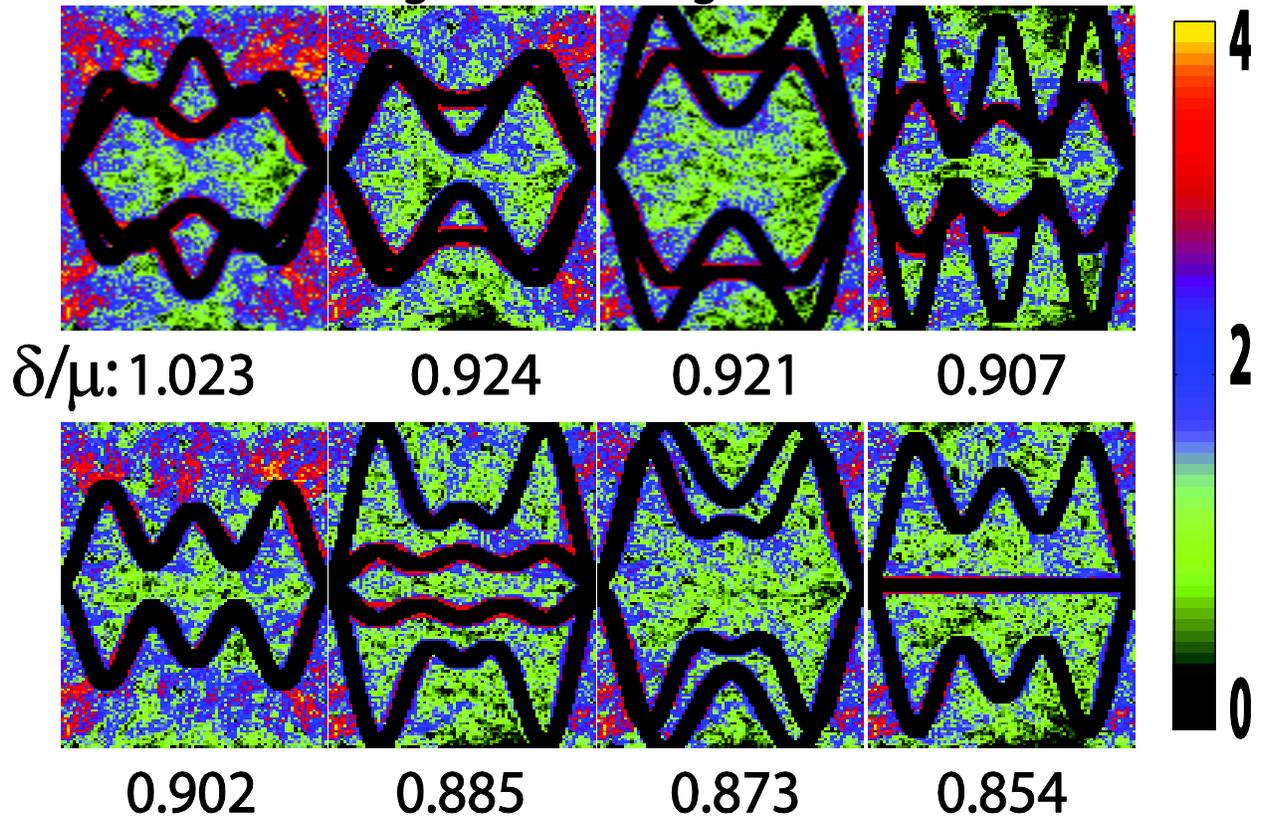

**Figure 2.** Shear stress distributions for eight different generations sampled from a simulation comprising a total of 100 generations. The topology optimization by genetic algorithm results in: (1) a target shear stress value of 1; (2) more uniform shear stress distribution around the target value.

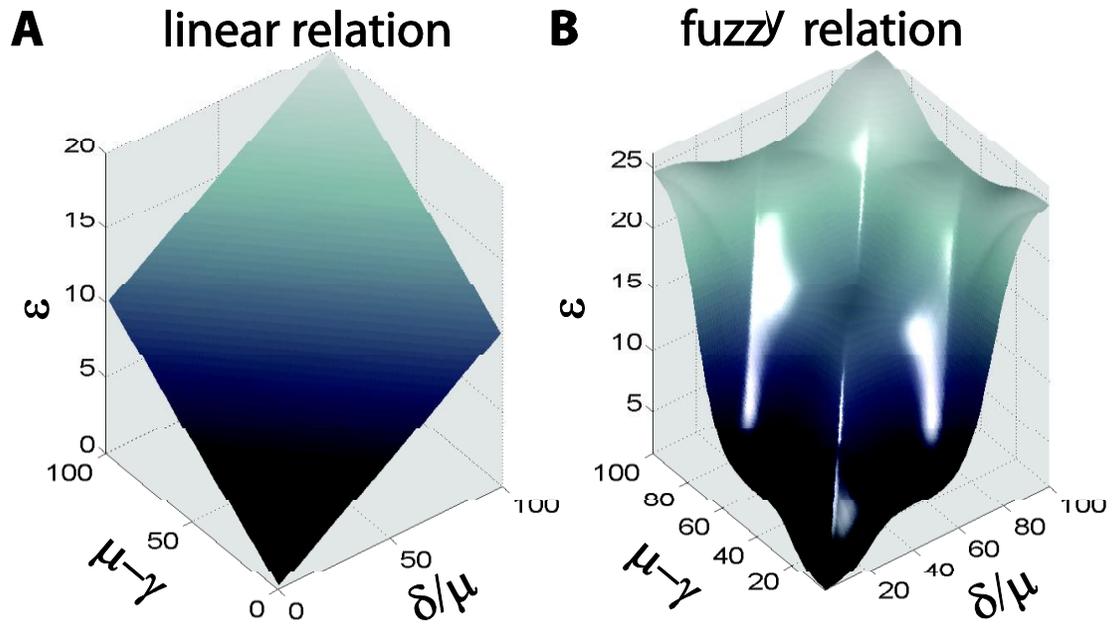

**Figure 3.** Comparison of error surfaces in the case of the (A) linear relation versus (B) fuzzy relation.

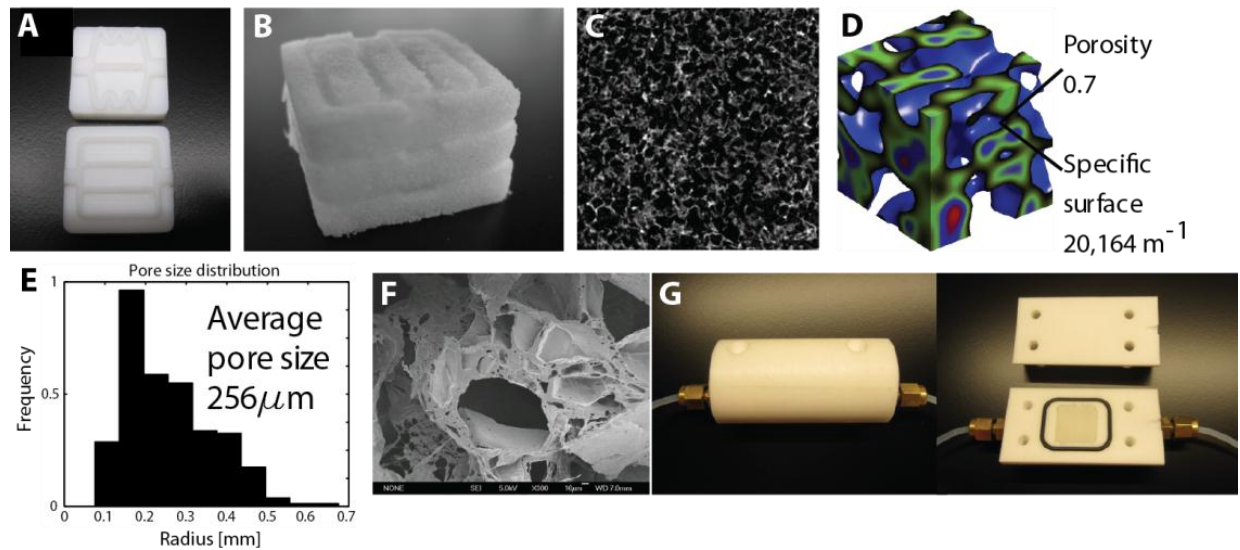

**Figure 4.** (A) Teflon molds used to cast sugar/PCL paste and fabricate 3D scaffolds. (B) The leached scaffolds are highly porous and can be stacked to achieve the desired thickness. (C) X-ray microcomputed tomographic (μ-CT) image revealing open porosity of scaffold, scanned at 13 μm resolution. (D) 3D rendering of μ-CT data using MATLAB to estimate porosity and specific surface. (E) Pore size distribution as determined by μ-CT data. (F) Scanning electron micrograph of a cross-section representative of the scaffold. (G) Teflon bioreactor shown assembled (left) and opened (right). The interior of the flow chamber is 2cm × 2 cm × 1 cm.

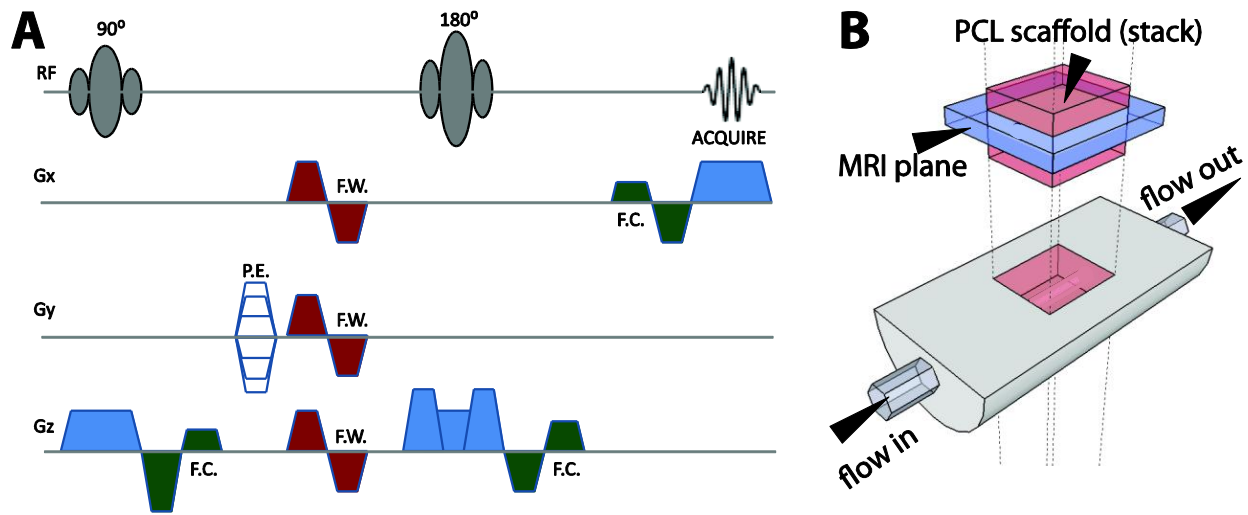

**Figure 5.** (A) Slice-selective, spin-echo MRI pulse sequence used for the flow measurements. "F.C." (green color) indicates the position of flow compensating gradients. "F.W." (red color) indicates bipolar flow weighting gradients. (B) Selection of the MRI slice in relation to the porous PCL scaffold. The slice position is selected to align with the inlet and outlet of the flow chamber.

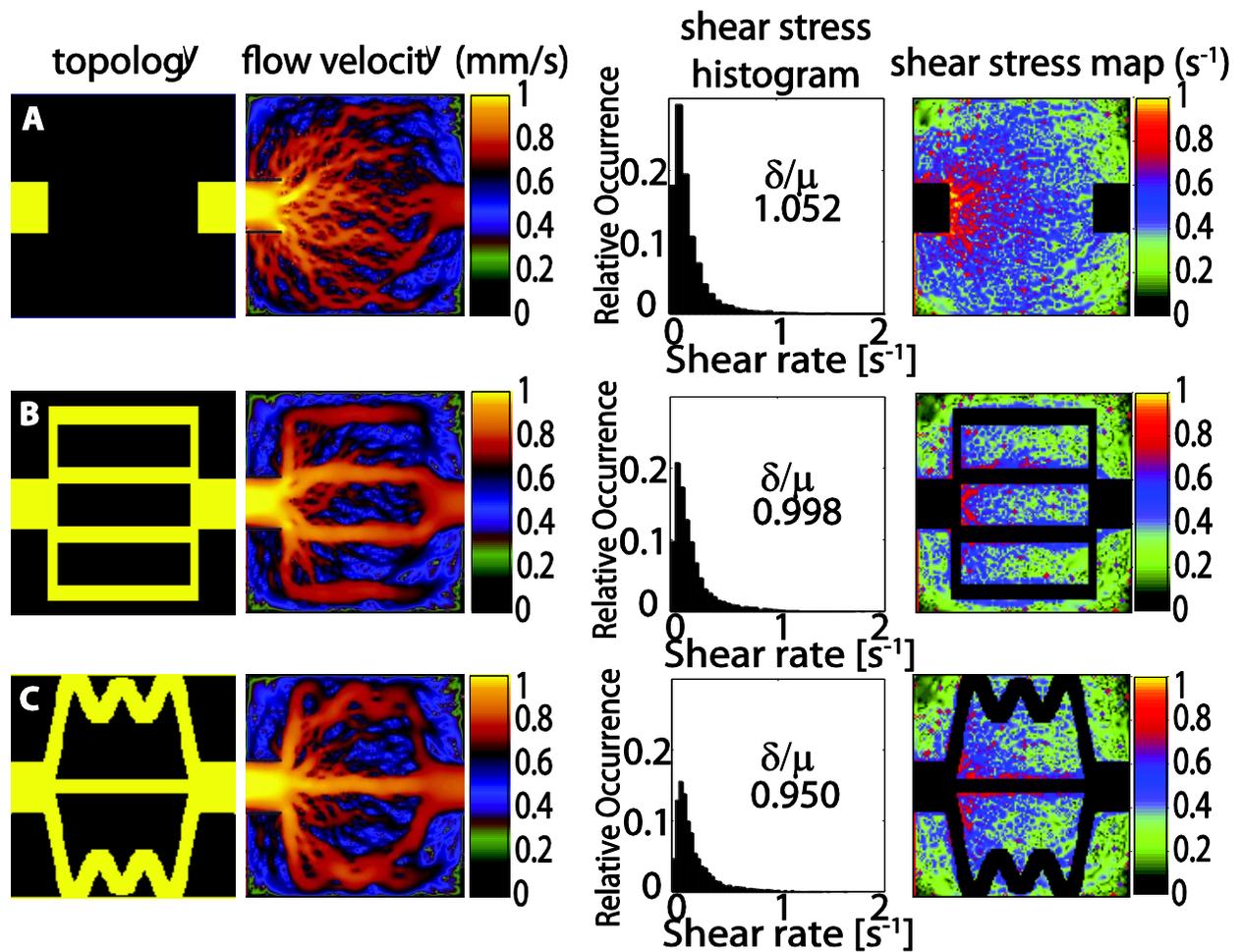

**Figure 6.** Computer simulation results for channel topologies and corresponding input velocities: (A) no channel, (B) manual design and (C) optimized channel design. The optimization is limited to a very small number of discrete cosine components to keep the design simple for machining purposes (see Figure 9C). A better optimization utilizing more discrete cosine components is shown in Figure 8. For display purposes, the logarithm of the flow velocity and shear stress maps is shown.

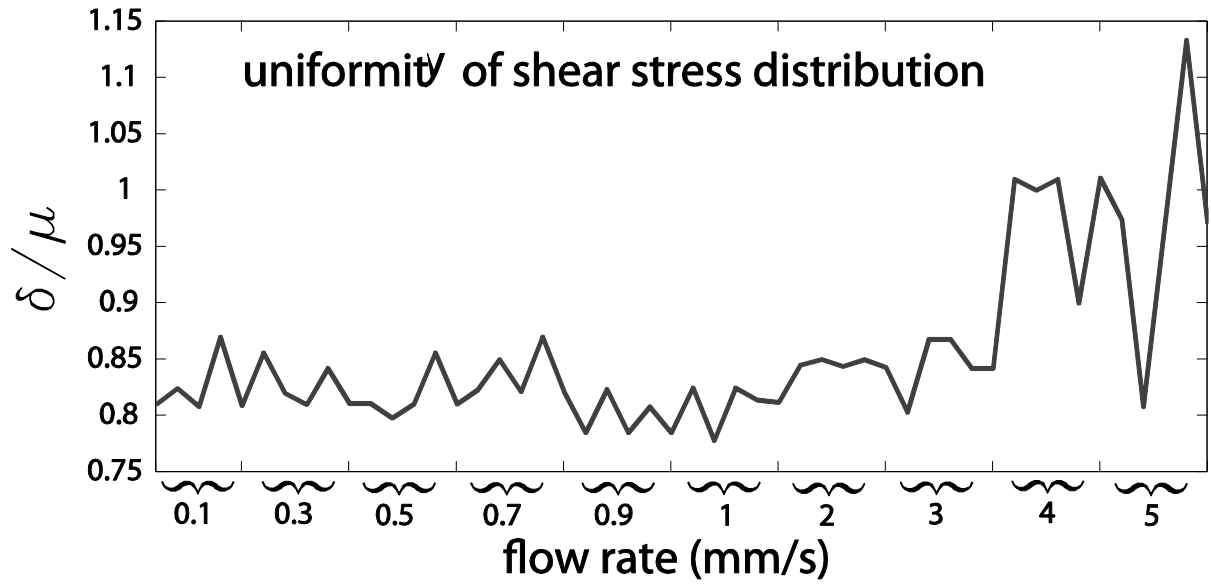

**Figure 7.** Uniformity of the shear stress distribution for different flow rates corresponding to input velocity values ranging from 0.1 to 5 mm/s.

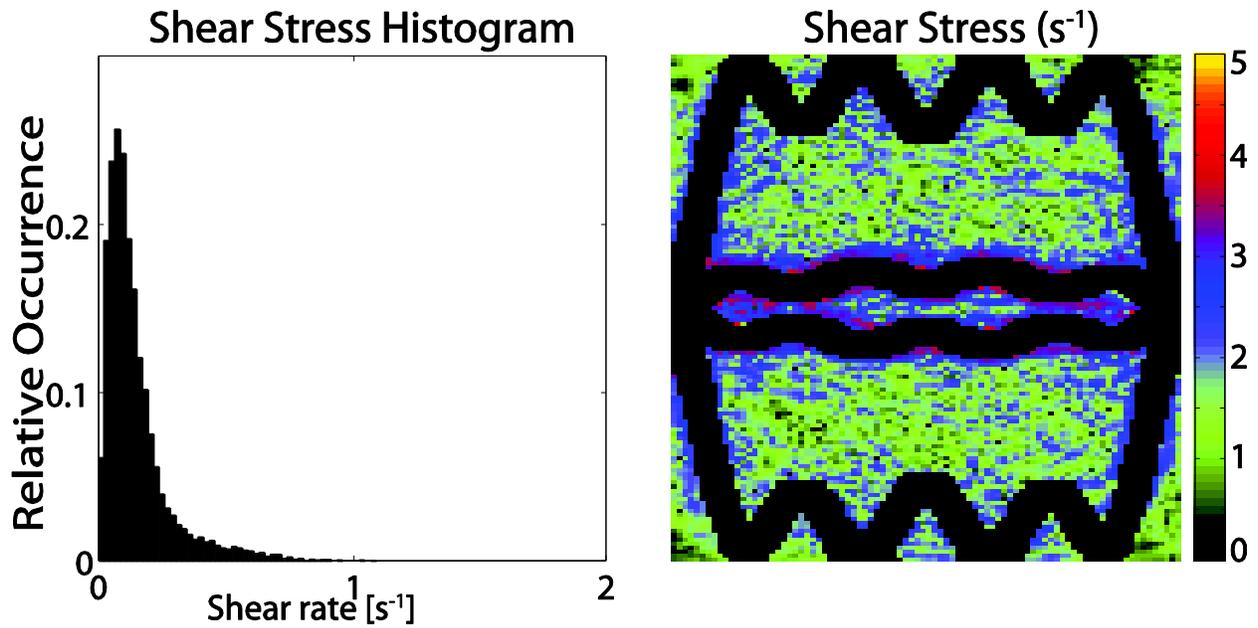

**Figure 8.** Shear stress histogram (left) and shear stress distribution (right) for an optimized topology. The flow rate was allowed to vary and more discrete cosine components were used than in the results of Figure 6. This resulted in improved performance.

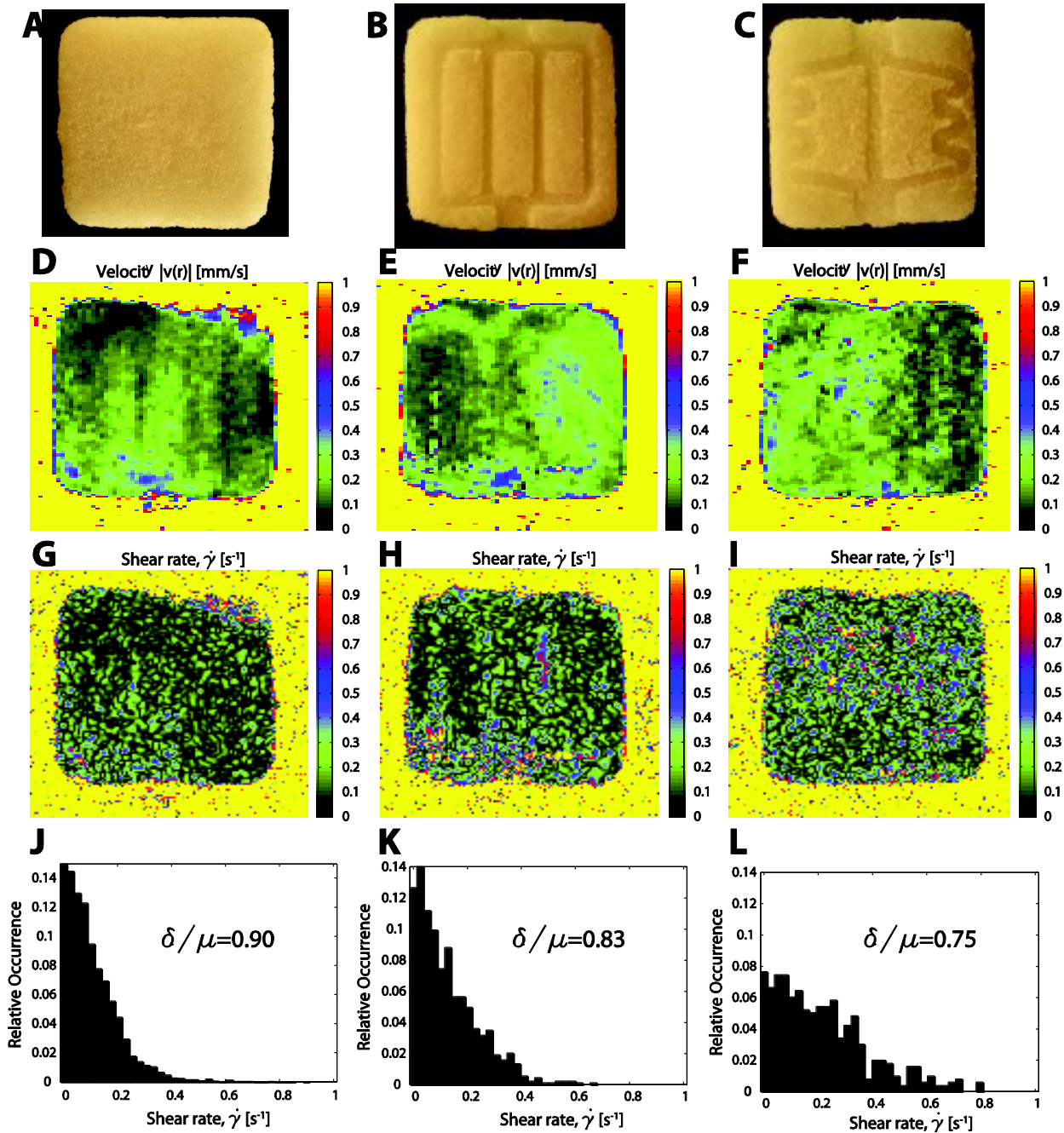

**Figure 9.** Comparison of MRI-derived shear rate distributions for water flowing in three different porous scaffolds: (A) no channel ($\delta/\mu \sim 0.90$), (B) manual design of the channel topology ($\delta/\mu \sim 0.83$) and (C) topology-optimized channel design ($\delta/\mu \sim 0.75$). In (A-C), a photograph of the 2cm × 2cm porous scaffold shows the channel topology used in the experiments. The flow rate in each scaffold was 5 mL/min at the inlet and resulted in flow velocities in the range of 0-1 mm/s inside the porous scaffold. Shear rate ($\dot{\gamma}$) distributions (G-L) inside the scaffold were derived from the appropriate flow map (D-F).